\documentclass[prd,notitlepage]{revtex4-1}
\usepackage{graphicx,amsmath,amsfonts,amssymb,rotating}
\usepackage[latin1]{inputenc}
\numberwithin{equation}{section}

\begin{document}

\def\eq#1{Eq. (\ref{eq:#1})}
\newcommand{\nc}{\newcommand}
\def\theequation{\thesection.\arabic{equation}}
\nc{\be}{\begin{equation}}
\nc{\ee}{\end{equation}}
\nc{\bea}{\begin{eqnarray}}
\nc{\eea}{\end{eqnarray}}
\nc{\barrayn}{\begin{eqnarray*}}
\nc{\earrayn}{\end{eqnarray*}}
\nc{\bcenter}{\begin{center}}
\nc{\ecenter}{\end{center}}
\nc{\ket}[1]{| #1 \rangle}
\nc{\bra}[1]{\langle #1 |}
\nc{\0}{\ket{0}}
\nc{\mc}{\mathcal}
\nc{\etal}{{\em et al}}
\nc{\GeV}{\mbox{GeV}}
\nc{\er}[1]{(\ref{eq:#1})}
\nc{\onehalf}{\frac{1}{2}}
\nc{\partialbar}{\bar{\partial}}
\nc{\psit}{\widetilde{\psi}}
\nc{\Tr}{\mbox{Tr}}
\nc{\tc}{\tilde c}
\nc{\tk}{\tilde K}
\nc{\tv}{\tilde V}
\nc{\CN}{{\mathcal N}}
\nc{\muphi}{\mu_{\phi}}

\title{Catastrophic Inflation}

\author{Sean Downes}
\author{Bhaskar Dutta}
\author{Kuver Sinha}
\affiliation{Department of Physics \& Astronomy, Texas A\&M University \protect\\
Texas A\&M University, College Station, TX 77843-4242, USA}

\begin{abstract}
We study inflection point inflation using Singularity Theory, which relates degenerate critical points of functions to their local behavior. This approach illuminates universal features of small-field models and gives analytic control over parametrized families of scalar potentials near inflationary solutions. The behavior of the scalar potential is tied to the number of physical input parameters, which determines a set of universality classes. Within these classes, we obtain universal scaling relations for density perturbations and the scale of inflation. In specific models, we show that the scale of supersymmetry breaking also possesses scaling behavior. We illustrate this general structure with a specific example: the Racetrack Inflation model in type IIB string theory, with the inflaton being the real part of the K\"ahler modulus, and the input parameters being flux dependent quantities that appear in the 4D $\mathcal{N}=1$ superpotential.
\end{abstract}

\preprint{MIFPA-11-21}

\maketitle

\section{Introduction}

A period of inflation in the early universe, dominated by vacuum energy, solves the flatness and horizon problems of big bang cosmology, in addition to diluting unwanted relics \cite{Komatsu:2010fb,Guth:1980zm,Linde:1981mu}. The vacuum energy is converted to radiation after inflation ends, which acts as the source of entropy for the universe. Most importantly, quantum de Sitter fluctuations of the inflaton leave an almost scale-invariant spectrum of perturbations on space-time, which leads to the formation of structure.

The minimum requirement for a functioning model of inflation is a scalar potential which generates sufficient e-foldings and is compatible with the observed values of the power spectrum $\mathcal{P}_{\mathcal{R}}(k)$ and the spectral index $n_{s}(k)$. A successful model should also satisfy the deeper need for a consistent embedding within particle physics and string theory. This has been an arena of much activity; for recent reviews, we refer to \cite{McAllister:2007bg,Baumann:2009ni}.

A sensible embedding within particle physics would ostensibly address the conceptual problems associated with trans-Planckian VEVs of the inflaton, which are hard to understand within the framework of effective field theory. Notable progress in this direction has been achieved \cite{Silverstein:2008sg,McAllister:2008hb}.

Another interesting and much-studied route has been the class of small-field models, especially models where inflation occurs near an inflection point of the inflaton potential. These offer a simple way to evade trans-Planckian VEVs $-$ the extreme flatness of the potential near the inflection point guarantees that enough e-foldings are generated while the field remains in the vicinity, provided the initial conditions are satisfied. Moreover, the scale of inflation can be low in these models, enabling a more direct connection with the scale of supersymmetry breaking. Inflection point models are common: D-brane inflation \cite{Baumann:2010sx,Kachru:2003sx,Chen:2009nk,Agarwal:2011wm}, K\"{a}hler modulus inflation~\cite{Conlon:2005jm,BlancoPillado:2004ns,BlancoPillado:2009nw}, MSSM inflation with soft terms~\cite{Allahverdi:2006iq}, and without soft terms~\cite{add} are all examples.

From a purely mathematical perspective, inflationary models that employ inflection points reduce to the study of functions whose first and second derivatives nearly vanish. These inflection or \textit{degenerate critical points} determine the qualitative behaviour of their functions.

In this paper, we study inflection point models using degenerate critical points. The appropriate language is Singularity Theory \cite{arnold1}, which relates degenerate critical points of functions to their behavior. Since the scalar potential governing inflation is a real function, we apply techniques from Catastrophe Theory. Catastrophe Theory studies critical points of real functions and therefore naturally describes inflection point inflation.

This mathematical language is useful in various ways. First, all models have a set of physical input parameters which determine the inflaton potential. Catastrophe Theory classifies the kinds of inflection points one can obtain by the number of control parameters. Second, for a given number of control parameters, Catastrophe Theory gives analytic control over the parameter space and therefore the regions where inflection points may be obtained. This is particularly useful for complicated potentials with many input parameters. Third, Catastrophe Theory grants universal scaling relations for physical observables such as energy scales and density perturbations.

These methods are applicable to virtually all models of inflection point inflation (with caveats spelled out later). In this paper we illustrate the general structure with a specific example: the well-studied Racetrack Inflation model in type IIB string theory \cite{BlancoPillado:2004ns}. The inflaton is the real part of the K\"ahler modulus, and the control parameters are flux dependent quantities that appear in the 4D $\mathcal{N}=1$ superpotential. There is a single variable and three independent control parameters, corresponding to an $A_{4}$ singularity or ``Swallowtail Catastrophe''. Parameter ranges which give acceptable inflation are obtained. 

The Hubble parameter during inflation is found to scale with the separation in field space of the point of inflation and a metastable vacuum with nearly vanishing cosmological constant. The vanishingly small vacuum energy was a numerical challenge for previous work \cite{Allahverdi:2009rm}, but is now readily found  and conceptually straight forward. 

For an appropriate choice of parameters, the vacuum may be widely separated from the point of inflection, leading to a separation of the scales of inflation and SUSY breaking \cite{Kallosh:2004yh}. Catastrophe Theory describes precisely how these scales separate.

The plan of the paper is as follows. In Section \ref{generalintro}, we introduce the machinery of Catastrophe Theory in a way that can be directly applied to inflection point inflationary models. In Section \ref{cusp}, we work out a simple toy model of inflation with two control parameters, corresponding to the ``Cusp Catastrophe''. In Section \ref{swallowtail}, we work out a more realistic example of inflation with three control parameters, corresponding to the ``Swallowtail Catastrophe''. We describe in detail the example of racetrack inflation, with particular focus on the scales of inflation and SUSY breaking. We conclude with a discussion.

\section{Machinery} \label{generalintro}
\subsection{Morse Theory and Inflation}\label{morse}
The study of inflation builds on the study of the scalar potential, which determines the stability of the physical system. Stability hinges on the existence of real critical points of the potential $-$ local extrema. Morse theory formalizes the familiar notion of stability. 

The local behavior of a function at a generic point is determined by its derivative there. Functions are typically increasing or decreasing. A point where the first derivative vanishes is called \textit{critical}. The local behavior at a critical point is determined by a nonvanishing second derivative. Stability of the system can be inferred from the the Hessian Matrix. If all the eigenvalues are positive, the system is stable. Negative eigenvalues correspond to unstable directions. In scalar field theory, these would be massive and tachyonic fields, respectively. Even better, small perturbations of such a potential don't change its stability properties. The function is said to be \textit{structurally stable} against perturbations.

When an eigenvalue of the Hessian vanishes, the function is no longer described by Morse theory. The system is marginally stable at the critical point, but structural stability is lost. For example, $x^3$ has a vanishing first and second derivative at the origin, but a small perturbation
\begin{equation}\label{function}x^3+ ax,\end{equation}
is classically stable for any $a<0$ (local extrema), but unstable for any $a>0$ (imaginary critical points). As $a$ approaches zero, the critical points degenerate. This non-Morse behavior of changing structural stability is the subject of Catastrophe Theory.

Notice that the structure of the function in \eqref{function} changes as $a$ passed through zero. Generally, functions will depend on many \textit{control parameters}, like $a$. We therefore speak of families of functions with a control parameter space, $\mathcal{C}$. In the context of effective field theory, this is the space of couplings. As familiar from Statistical Mechanics, couplings can be tuned to yield Critical Phenomena, where scale invariant thermal (or quantum) fluctuations dominate the physical system. This occurs when critical points of the function describing the system degenerate. For inflation, these fluctuations manifest themselves as the nearly scale invariant spectrum of density perturbations.

Now, a sufficient condition for inflation is that the slow roll parameters nearly vanish. Small-field inflation imposes a stricter version of this condition, namely,
\begin{equation}\label{slowroll} \Big|\frac{\partial V}{\partial\phi}\Big|_{\phi=\phi_0}\ll1,\quad \Big|\frac{\partial^2V}{\partial\phi^2}\Big|_{\phi=\phi_0}\ll1 \;.\end{equation}

Thus, small-field inflation occurs near a degenerate critical point. 

\subsection{Catastrophe Theory and Inflation}

Catastrophe Theory explains how the stability structure of Morse functions change. Such change occurs when critical points degenerate, which is also a condition for Inflation. The observation, illustrated by \eqref{function}, results in a domain structure for $\mathcal{C}$. This fact will yield important relationships between this domain structure, inflation and the physics in which inflationary models are embedded. This section is devoted to generalities. We will discuss two explicit examples of how such domain structures form in sections \ref{cusp} and \ref{swallowtail}.

For clarity, we begin with a single field. Let $V(x)$ be the potential associated with a scalar field $x$. $V$ is a real function which also depends on a number of couplings. Physically, we are often interested in the whole space of possible couplings, $\mathcal{C}(V)$. 

As discussed above, the critical points $\beta_{i}$ determine the structure of a function. To illuminate this fact, one may write the derivative of $V$ as
\begin{equation}\label{heart}V^{\prime}(x)=v(x)\prod_{i}(x-\beta_{i}),\end{equation}
where $v(x)$ never vanishes. Clearly, the critical points $\beta_i$ will change with the control parameters in $\mathcal{C}(V)$. Since $V$ is a real function, the $\beta_{i}$ may appear in complex conjugate pairs. Real roots correspond to local extrema. As discussed above, the stability structure of $V$ changes (minima disappear) as $\beta_{i}$ pass between real and complex values. This leads to the observation of singular importance in Catastrophe Theory: such transitions necessarily occur when the critical points are identical $-$ they degenerate. 

A generic point in $\mathcal{C}(V)$ corresponds to distinct $\beta_{i}$, and thus a Morse function. For an $n$-dimensional parameter space, however, there exist subsets of couplings where the $\beta_i$ degenerate. For example, if two of the critical points take the same value $\alpha$,
$$\beta_1=\beta_2=\alpha,$$
this corresponds to a single constraint between the couplings in $\mathcal{C}(V)$. This single constraint corresponds to a hypersurface in $\mathcal{C}(V)$ of codimension one $-$ a domain wall. Roots transition between real and complex values only as the couplings jump across the wall. These walls divide the space of control parameters into domains where $V$ has different stability structures.

For a generic point on the domain wall, $V$ has a doubly degenerate critical point, $\alpha$ and may be expressed locally by its Taylor series,
\begin{equation}\label{taylor}V(x)\approx V(\alpha) + \frac{\lambda_{3}}{3}(x-\alpha)^3 + \mathcal{O}[(x-\alpha)^4].\end{equation}
The linear and quadratic terms vanish; $V$ is non-Morse. More importantly, the slow roll conditions for small-field inflation \eqref{slowroll} are satisfied in, and only in, the vicinity of a domain wall.

Higher degeneracy among the critical points can be described similarly. An $m$-fold degenerate critical point,
$$\beta_{1} = \hdots = \beta_{m} = \alpha,$$
corresponds to a $(n-m)$-dimension hypersurface in $\mathcal{C}(V)$. This process terminates at $m=n$, when the maximum possible critical points degenerate. This corresponds to a point $\mathcal{C}(V)$, $c_{\diamond}$. The function, $V(x)$, associated with the couplings at $c_{\diamond}$ is called the \textit{germ} of the Catastrophe. This ``domain complex'' will have important implications for the effective field theory. 





There is a technical point worth mentioning. The mathematical space of control parameters, say $\bar{\mathcal{C}}$, is $n$ dimensional, but $\mathcal{C}$ may have lower dimension. This just means that the physical parameter space is less than the most general allowed by mathematics. This is common, particularly given some imposed symmetry. For simplicity we shall ignore these cases and assume\footnote{A simple example of a nontrivial restriction of the parameter space would be two widely separated degenerate critical points. While they do descend from the unfolding of a critical point of higher degeneracy, it may be physically simpler to consider them as isolated Catastrophes with their own parameter spaces with some appropriate matching conditions. This is reasonable because Catastrophe theory involves the local behavior of functions. Indeed, the intuition gained from ``gluing'' is helpful when considering more complex Catastrophes, such as those that involve multiple variables.} $\mathcal{C}=\bar{\mathcal{C}}$, and discuss explicit exclusions of parameter space where appropriate.
 


 
Inflation occurs in the vicinity of a domain wall. Deviations of the couplings away from some point on the domain wall can be encoded in a nonvanishing $\lambda_{1}$. This parametrizes the ``unfolding'' of the degenerate critical point. This means that \eqref{taylor} becomes,
\begin{equation}\label{taylor2}V(x)\approx V(\alpha) + \lambda_{1}(x-\alpha) + \frac{\lambda_{3}}{3}(x-\alpha)^3 + \mathcal{O}[(x-\alpha)^4].\end{equation}
Here $\alpha$ is still the point of inflection: $V^{\prime\prime}(\alpha)=0$. 
Notice that the slow roll parameter $\epsilon = \frac{1}{2}\lambda_{1}^2$. In the slow roll approximation, one may compute the number of e-foldings associated with \eqref{taylor2},
\begin{equation}\label{whos}N_{e} = \int_{0}^{x_{\rm end}}dx\frac{V}{V^{\prime}} = \int_{0}^{x_{\rm end}}\frac{dx}{\lambda_{3}x^2 + \lambda_1},\end{equation}
which, to a very good approximation, is
\begin{equation}\label{efolds}N_{e} = \frac{\pi}{2\sqrt{\lambda_3\lambda_1}}.\end{equation}
To ensure the proper numbers of e-foldings, $\lambda_{1}$ must be quite small. This leads us to our first fundamental result: $\lambda_1$ is essentially model independent and is degenerate with the number of e-foldings \footnote[2]{This is certainly true in the slow roll approximation, Eq. (\ref{whos}). Interestingly, this remains true when a broader set of initial conditions is considered. This is be explored in future work.}. This is important, as $\lambda_{3}$ deviates only infinitesimally from the domain wall value where $\lambda_1=0$. As we will demonstrate with an explicit example in section \ref{swallowtail}, this distinction is physically unimportant. Thus, $\lambda_{3}$ depends solely on a subset of couplings which parametrize the domain wall. As we shall also see, $\lambda_{3}$ can be computed explicitly once the type of Catastrophe is chosen. Taken together, these facts yield our second fundamental result: all physical observables connected with the inflation sector are determined by the domain wall parameters, and may therefore be classified by Catastrophes. We now turn to the classification of these Catastrophes.
\subsection{ADE Classification of Inflaton Potentials}
We now present Arnold's classification of Catastrophes relevant for the scalar potential. It relies one two numbers: dimensionality of $\mathcal{C}(V)$ and number of possible vanishing eigenvalues of Hessian. There are also technically important minus signs. 

The \textit{Splitting Lemma} is an important result for this classification. It proves the existence of a ``good'' coordinate patch on the target space $\Sigma$ where Morse and non-Morse behavior can split into different variables. Mathematically, this removes any constraint on the dimensionality of $\Sigma$ and allows us to focus only on the number of vanishing eigenvalues of the Hessian matrix. Physically, it means that we can always find a field redefinition which splits all scalar fields into an ``inflaton'' (or the relevant multifield configuration) and ``spectators''. Of course, one must always ensure that spectators are stabilized during inflation.

Arnold's work involved complex functions. Since we are interested in the scalar potential, we extract what is relevant for real functions. Arnold demonstrated that the Catastrophe domain structures are linked to symmetries of the space of Taylor polynomials (Jets) at a critical point of $V$. Specifically, this space at a degenerate critical point is invariant under the action of some maximal Lie Group $\mathcal{G}$. The larger the symmetry group, the more complicated the behavior, and the higher order the Taylor polynomial. For explicit details of the group action see the appendix.

Simply put, if there is only a single vanishing eigenvalue of the Hessian, $\mathcal{G}$ must be the group $A_{k}$, where rank $k$ is the dimension of the control parameter space. If there may be two vanishing eigenvalues, the group may also be $D_{k}$ or $E_{k}$. Technical issues arise if there may be more than two vanishing eigenvalues. Given such a case for $\dim\mathcal{C}>6$, for instance, the critical point is associated with a one parameter family of germs. $\mathcal{G}$ may no longer be simple, but may pick up a $\mathsf{U}(1)$ factor, for instance. While elegant, this is far beyond the scope of our work. For proofs and a more detailed discussion, see Arnold \cite{arnold1}.

These considerations are deep, but the results are simple. For all relevant applications, the scalar potential $V$ is mapped into a finite polynomial which depends on the Catastrophe Structure of $V$. Usually, this map is just the relevant Taylor polynomial.

In what follows, $x,y$ are fields parametrizing the some target space $\Sigma$ and other parameters belong to $\mathcal{C}$. 
\begin{eqnarray}
A_{\pm k}&:\;& (\pm)^{k}x^{k+1} + \sum_{m=1}^{k-1}a_{m}x^{m},\\
D_{\pm k}&:\;& (\pm)^{k} xy^2\pm x^{2k-1} + \sum_{m=1}^{k-3}a_{m}x^m + c_1 y + c_2 y^2,\\
E_{\pm 6}&:\;& \pm(x^4 + y^3) + a x^2y + b x^2 + c xy + d x  + f y,\\
E_{7}&:\;& y^3 + yx^4 + \sum_{m=1}^{4}a_m x^m + b y + c xy,\\
E_{8}&:\;& x^5 + y^3+y\sum_{m=0}^{3}a_{m}x^m + \sum_{m=1}^{3}c_{m}x^m.
\end{eqnarray}
In this representation, the origin of $\mathcal{C}$ corresponds to $c_{\diamond}$.

Applying this classification to a smooth potential is straight forward. If only a single eigenvalue of the Hessian may vanish, then the Catastrophe depends on the number of couplings. For $n$ couplings \textit{exclusive} of a constant term, we have $A_{n+2}$ behavior. When two eigenvalues may vanish, $D_{k}$ behavior is typically associated with $\dim\mathcal{C}(V)=k$. Of course for $k=5,6,7$, one must be wary of exceptional germs. In this work we focus primarily on the single field case, leaving the more complicated multifield parameter spaces for future work.


\section{$A_{3}$: The Cusp Catastrophe} \label{cusp}
\subsection{$A_{3}$ Domain Structure}
A potential with $A_{3}$ behavior is simplest workable model of inflation. A triply degenerate critical point can unfold to leave a single doubly degenerate critical point for inflation. The remaining root is a local minimum - the vacuum.  This class of potential exhibits the majority of relevant behaviors. In this section we investigate we build the domain structure of $A_{3}$ and examine the physical details.

\begin{figure}[!h]\centering
\includegraphics[width=4in]{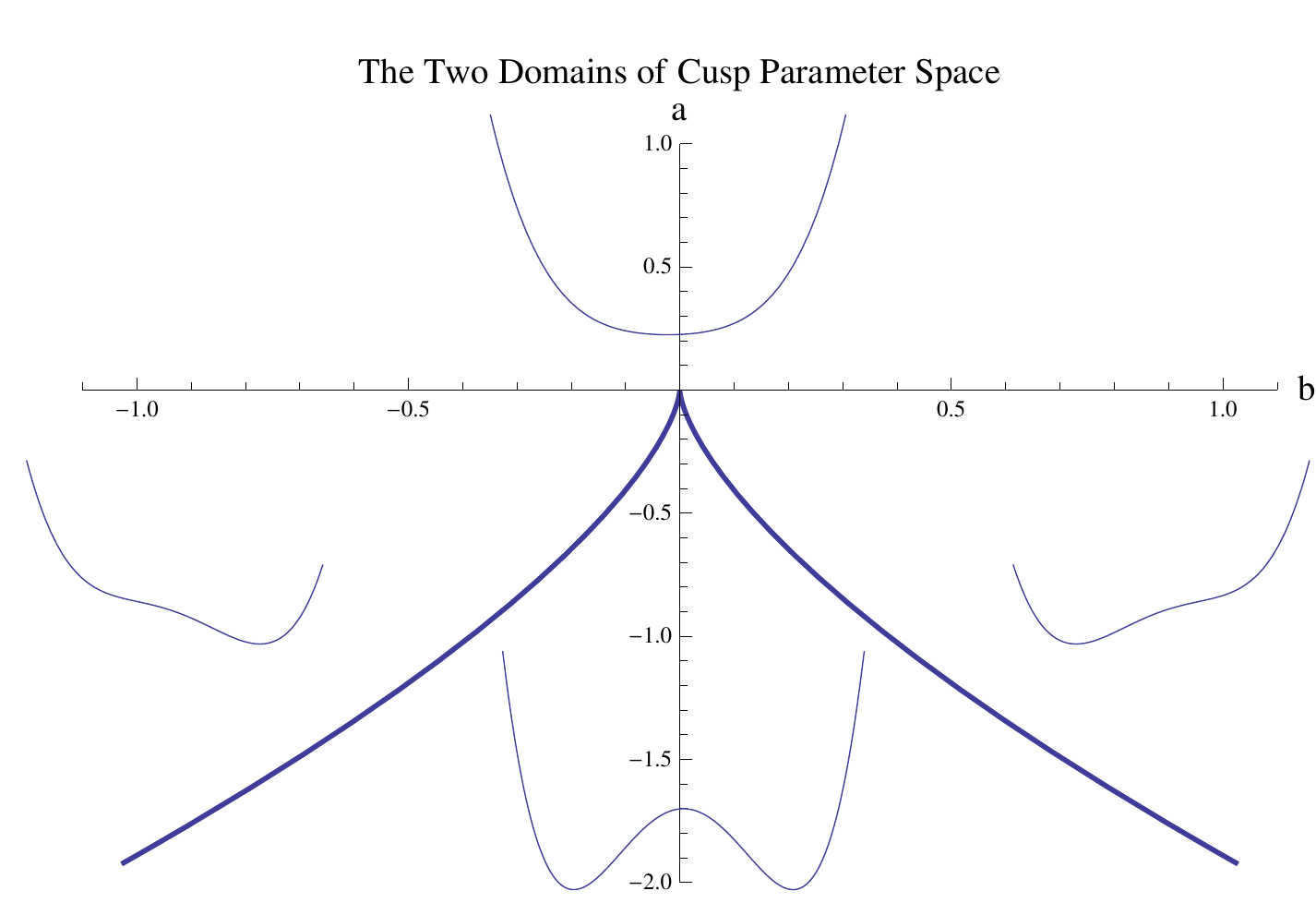}\caption{A divide between two different regions of Cusp parameter space.}
\label{cuspfig}
\end{figure}

The most general, local form for a potential with with $A_3$ behavior is
\begin{equation}\label{cusppot} V(x) = \frac{\lambda}{4}x^4 + \frac{1}{2}ax^2 + bx + c \;. \end{equation}
For stability, we require $\lambda>0$. Demanding $V$ support both inflation at a degenerate critical point as well as a stable vacuum elsewhere requires all three critical points of $V$ to be real, and two to be degenerate. Inflation occurs at the latter, $x=\alpha$, and ends with the field rolling into the final vacuum at $x=\beta$. There the inflaton decays and reheating begins. In short,
\begin{equation}\label{dcusppot}\frac{dV}{dx} = \lambda (x-\alpha)^2(x- \beta)\;. \end{equation}
We now illustrate some features of this simple model. Taking a derivative of Eq.~\ref{cusppot}, and matching coefficients in Eq.~\ref{dcusppot}, we find
\begin{equation}\label{edit1}
2\alpha + \beta = 0,
\end{equation}
and the relations
\begin{equation}
\quad \alpha^2 + 2\alpha\beta = a/\lambda,\quad -\alpha^2\beta = b/\lambda \,\,.
\end{equation}
Eliminating $\beta$, one finds,

\begin{equation}\label{ab} 
a/\lambda = -3\alpha^2,\quad b/\lambda = 2\alpha^3\,\,.
\end{equation}
Eliminating $\alpha$, we have the constraint
\begin{equation}\label{cuspeqn}
\Big(\frac{a}{3\lambda}\Big)^3 + \Big(\frac{b}{2\lambda}\Big)^2 = 0\,\,.
\end{equation}
Eq.~\ref{cuspeqn} is the condition that $V$ must satisfy in order to have an inflection point. In other words, Eq.~\ref{ab} defines a domain wall in parameter space. The name of the Cusp Catastrophe originates from the shape of the domain wall, as seen in Figure \ref{cuspfig} (with $\lambda=1$).

It is instructive to study the nature of the parameter space away from the domain wall, at a generic point of the two-dimensional\footnote{Mathematically, the additive constant in all potential functions adds a trivial dimension to the parameter space. As such, it is traditionally set to zero. Physically, it is intimately related to the cosmological constant. In view of this, we shall carry it around explicitly, but not treat it as parameter of $V$.} parameter space of $V$. 
As Figure \ref{cuspfig} shows, the cusp splits the parameter space into two domains. Away from the domain wall, the three critical points of $V$ are distinct. In the upper domain, two of the three roots are imaginary; there is only a single minimum. Below, they are all real. Notice that $b\rightarrow-b$ has the same effect as a parity transformation, $x\rightarrow -x$. 

The physics here is stable. Near any value of $a$, there are two disconnected neighborhoods of $b$ which support inflation. Perturbations of the scalar potential simply push these neighborhoods around. Inflation is possible as long as the physical parameter space includes the domain wall. This structure appears in the analysis of higher catastrophes. We shall see it embedded in the domain structure of $A_{4}$ in section \ref{swallowtail}. This implies that higher order corrections cannot exclude inflation unless they exclude a domain wall. This is the spirit of Catastrophe Theory. 


Apart from the neat description of parameter space, there are concrete, analytic results for inflationary physics. The separation of roots, $(\beta-\alpha)$, parametrizes many physical observables in a simple way, which we now investigate.
\subsection{$A_{3}$ Inflation}
Very close to the inflection point, $\alpha$, we expect to find cubic behavior of the potential. To that end, let $y=x-\alpha$. Substituting into Eq.~\eqref{cusppot}, together with Eq.~\eqref{edit1} yields,
\begin{equation}\label{newcu}V(y) = \frac{1}{4}y^4 + (\beta-\alpha) y^3 + \lambda_1 y+(\frac{\alpha^4}{12} + c).\end{equation}
Here, as in \eqref{taylor2}, $\lambda_1$ represents a deviation in $\mathcal{C}$ from the domain wall. It is the slow roll parameter, $\epsilon = \lambda_1^2/2.$ Eq.~(\ref{efolds}) gives number of e-foldings of inflation,
$$N_{e} = \frac{\pi}{2\sqrt{\lambda_1(\beta-\alpha)}}.$$

%
%


%
%


The Hubble parameter during inflation is related to the scale of inflation, which we may compute from \eqref{newcu}
$$H^2 = V(\alpha)=\frac{1}{12}(\beta-\alpha)^4 + V(\beta).$$
Here $V(\beta)$ is the final vacuum energy of the inflaton field. Finally, we come to the density perturbations. Integrating in Eq.~\eqref{whos} instead to $x_{\rm COBE}$ where COBE perturbations leave the horizon gives $N_{\rm COBE}$, which is used to normalize the computed density perturbations to observations. Replacing $x_{\rm end}$ and $x_{\rm COBE}$ with $N_{e}$ and $N_{\rm COBE}$, one finds that the density perturbations,
$$\Delta_{\mathcal{R}}^{2} = \frac{1}{4\pi^2}\Big(\frac{H^2}{dx/d N_e}\Big)^2\Big|_{x=x_{\rm COBE}},$$
may be written as \cite{Allahverdi:2009rm}
$$\Delta_{\mathcal{R}}^{2} = \frac{V(\alpha)}{12\pi^2}(\beta-\alpha)^2 N_{e}^{4}.$$

Immediately we see that density perturbations scale parametrically with $(\beta-\alpha)$,
$$\Delta_{\mathcal{R}}^{2} = \frac{V(\alpha)}{12\pi^2}N_e^4 (\beta-\alpha)^2.$$
If, as in our universe, $V(\beta)\approx 0$, we have the simple result,
$$\Delta_{\mathcal{R}}^{2} = V_0\frac{N_e^4}{144\pi^2}(\beta-\alpha)^6.$$
%

The physical observables discussed scale with $\beta-\alpha$. As one may read from Eq.~\eqref{ab}, $a$ and $b$ vanish when $\alpha =\beta$. The energy difference between inflation and the vacuum also vanishes. This is the consequence of $V$ having a triply degenerate critical point at $x=0$, that is,
$$V = \frac{1}{4}x^4 + c.$$
Notice that $N_{e}$ and $\Delta_{\mathcal{R}}^2$ must be recalculated for this case. 


We close this section with Gilmore's \cite{gilmore} remark about the moniker ``Catastrophe Theory''. Had we taken the dual catastrophe, $A_{-3}$ with germ $-x^4/4$, rather than $A_{3}$, the same arguments would result in the function whose graph is reflected about the $x$ axis. The three extrema phase would include a single, meta-stable minimum between two maxima.  Upon passing through the anti-Cusp boundary, the system would lose metastability, tending towards a violent runaway. Catastrophe!

\section{$A_{4}$: The Swallowtail Catastrophe} \label{swallowtail}


Realistic models, like flux compactifications, typically involve more structure than an $A_{3}$ potential affords. Indeed, the KKLT mechanism often includes a barrier to decompactification associated with SUSY breaking. This structure is only slightly more complex, and can be found in scalar potentials with $A_4$ structure. In this section we describe inflation for $A_{4}$ potentials along the lines of Section \ref{cusp}. We then apply these results to the study of K\"{a}hler moduli inflation in Type IIB String Theory.

\subsection{$A_{4}$ Domain Structure} \label{swallowparameter}
%
%


The most general, local form for a potential with $A_{4}$ behavior is
\begin{equation}\label{st}
V(x) = \frac{\lambda}{5}x^5 + \frac{a}{3}x^3 + \frac{b}{2}x^2 + cx + d\,\,.
\end{equation}
%

Here $\lambda$ may be positive or negative. Alternatively, one can take $\lambda>0$ and connect to other models by a parity transformation, $x\rightarrow -x$. As with $A_{3}$, the parameter space is divided into domains with different numbers of local minima. Again, we investigate inflation near doubly degenerate critical points. The same requirements of inflation and a local minimum forces the fourth critical point, $\gamma$, to be real, thus
\begin{equation}\label{dst}
\frac{dV}{dx} = \lambda(x-\alpha)^2(x-\beta)(x-\gamma)\,\,.
\end{equation}
Taking a derivative of Eq.~\eqref{st} and matching coefficients with Eq.~\eqref{dst}, we find
\begin{equation}\label{stcon}
2\alpha + \beta + \gamma = 0,
\end{equation}
and
\begin{equation}\label{dj}
a/\lambda = -(3\alpha^2 + 2\alpha\beta + \beta^2),\quad b/\lambda= 2\alpha (\alpha + \beta)^2,\quad c/\lambda = -\alpha^2\beta(2\alpha+\beta)\,\,.
\end{equation}
Like $A_{3}$ potentials, the behavior of $V$ depends on the sign of $a$. For $a\ge0$ parameter space has two domains: zero or one minimum. A domain with two minima opens only when $a<0$. Demanding inflation and a metastable vacuum restricts are attention to the latter. Precise analysis of these regions can be accomplished with a trick from \cite{gilmore}.

\begin{figure}[!h]\centering
\includegraphics[width=4.5in]{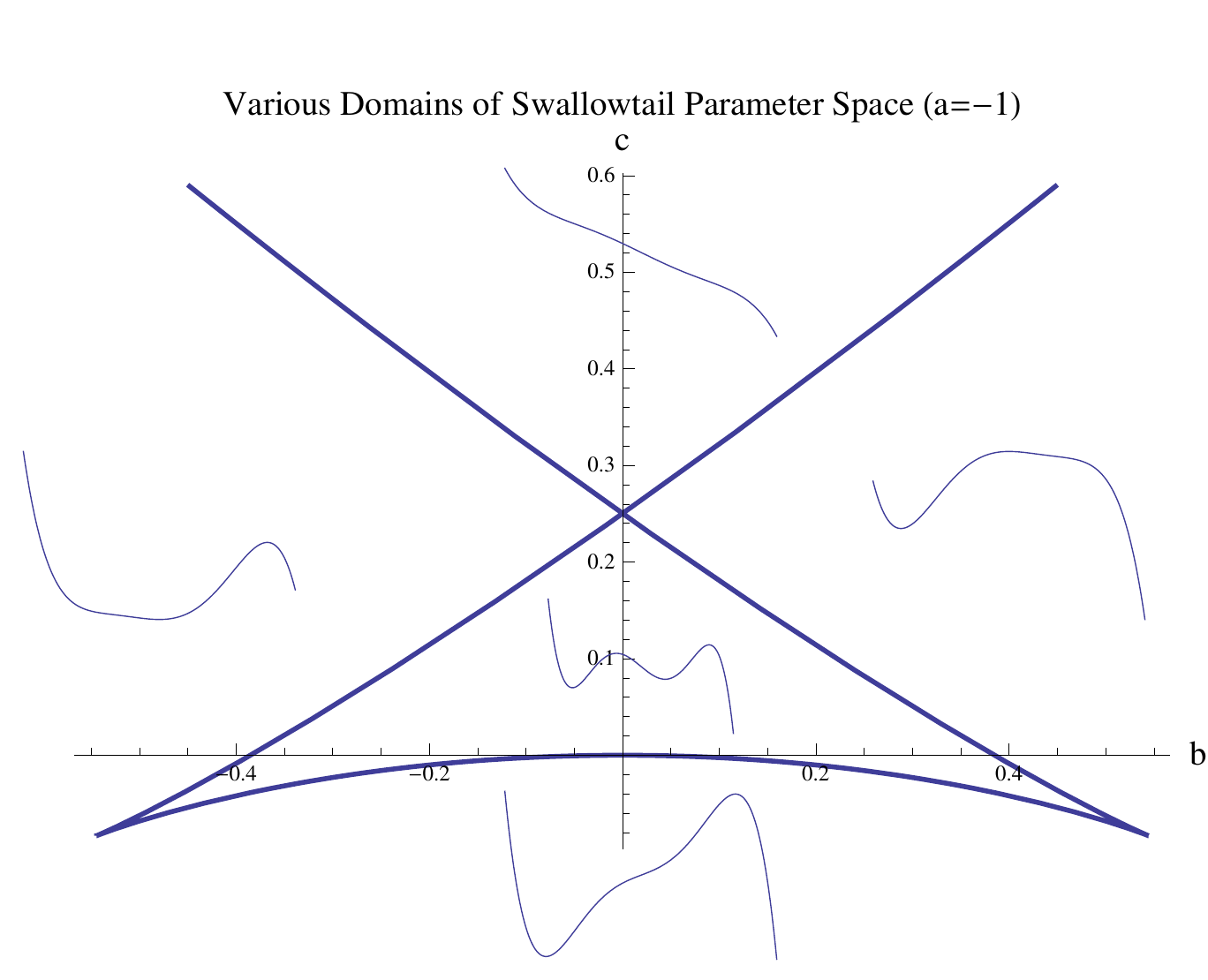}\caption{A divide between two different regions of Swallowtail parameter space, for fixed $a<0$.}
\label{swall}
\end{figure}

Under the scaling $x\rightarrow \zeta x$, demanding $V\rightarrow \zeta^5$, we find 
\begin{equation}
a\rightarrow \zeta^2 a,\quad b\rightarrow \zeta^3 b,\quad c\rightarrow \zeta^4 c \,\,.
\end{equation}
It suffices, then, to consider the cases $a=0,\pm 1$ to understand the full behavior of the parameter space. The rest may be accessed via scaling. We shall consider the case of interest, $a=-1$. Eliminating $\beta$ in Eq.~\eqref{dj}, we have the equations for the domain wall:
\begin{equation}\label{ndj}
\frac{b/\lambda}{|a/\lambda|^{3/2}} = -4\alpha^3 + 2\alpha,\quad \frac{c/\lambda}{|a/\lambda|^{2}} = 3\alpha^4 -\alpha^2 \,\,.
\end{equation}
$a$ may be arbitrarily negative. To simplify \eqref{ndj}, one may simply redefine parameters to effectively take $\lambda=1$.  Fixing $\zeta$, we have a cross section of the diagram, parametrized by the location of the degenerate root $\alpha$. This is illustrated in Fig~\ref{swall}.

\begin{figure}[!h]\centering
\includegraphics[width=5in]{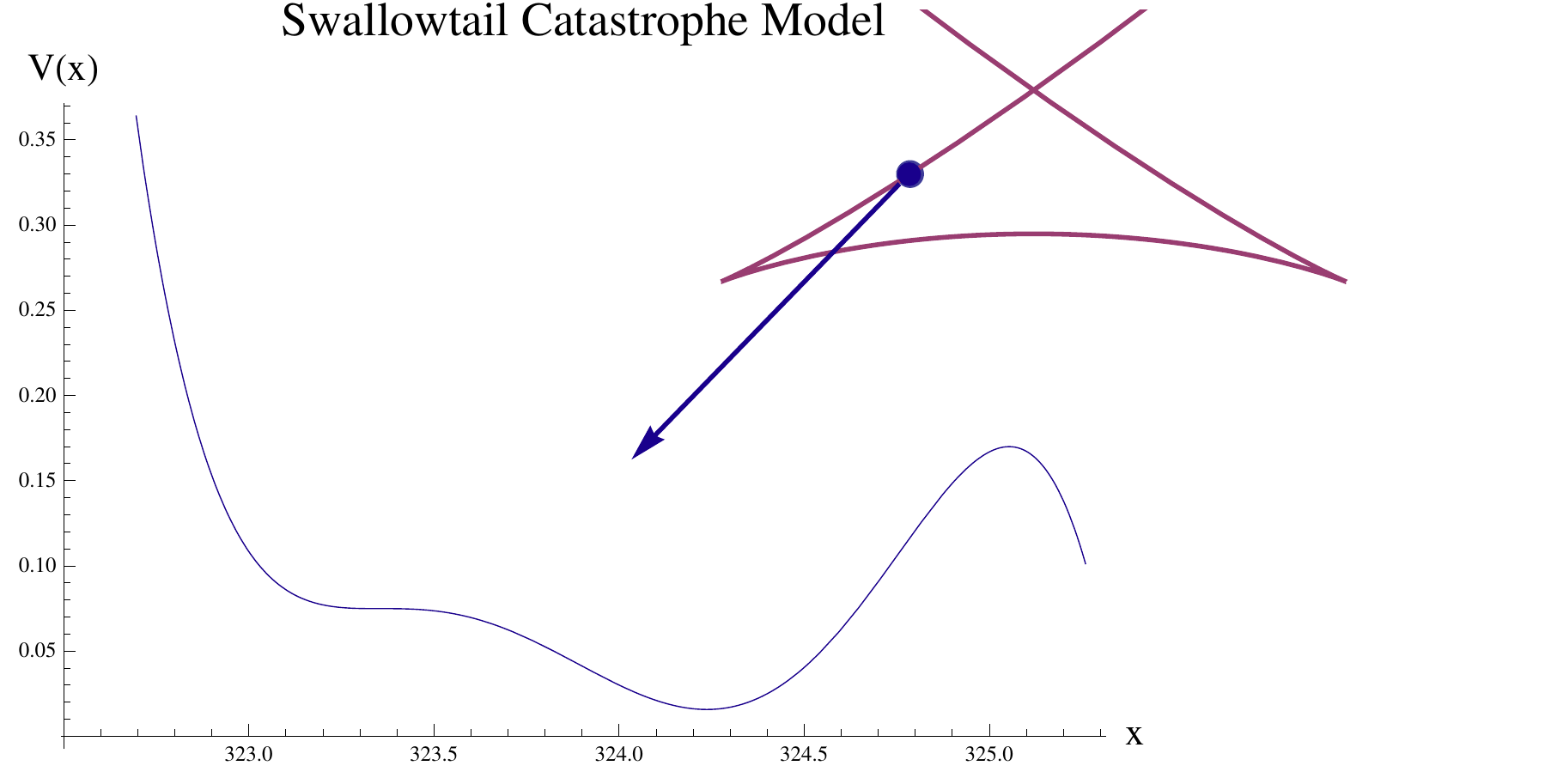}\caption{The Swallowtail potential has a doubly degenerate critical point at generic points on the domain wall of parameter space. This is illustrated for the $a=-1$ cross section of the three dimensional space. Note the similarity of structure to the typical Racetrack potential.}
\label{newswall}
\end{figure}

At a generic point on this curve, $V$ has an inflection point suitable for inflation. This is illustrated in Fig~\ref{newswall}. The two corners correspond to triply degenerate critical points. One can easily check that the left-hand corner behaves like the $A_{3}$ Catastrophe. The right-hand corner has $A_{-3}$ behavior. At the top vertex corresponds to two, separated, doubly degenerate critical points: an $A_{2}$ pair. The triangular, two minima region collapses when $a=0$. Here, four extrema degenerate into a single critical point. At this point we find the quartically degenerate critical point, the $A_4$ germ. For $a>0$, a parabola acts a domain wall between an unstable region and one with a single minima. We shall not consider this case further, but move on to physical considerations.


%


\subsection{$A_{4}$ Inflation} \label{swallowinflation}
We now investigate the properties of $V$ near a domain wall. As with \eqref{newcu}, let $y=x-\alpha$, where $\alpha$ is again the inflection point.
\begin{equation}\label{fold}V(x)\approx V_0\big(\frac{\lambda_3}{3}(x-\alpha)^3 + \lambda_1 (x-\alpha)\big) + V(\alpha).\end{equation}
The ``scaling coefficient'' of the $x^3$ behavior near the point of inflection is given by
$$\lambda_3 = -(\alpha-\beta)(\alpha-\gamma).$$
An inflection point suitable for inflation exists when $\lambda_1=0$, which is precisely true if Eq.~\eqref{dst} holds. This expression can be simplified with Eq.~\eqref{stcon}. Applying this constraint is tantamount shifting the entire potential by a fixed $\big<x\big>$. This simplified the understanding of Swallowtail behavior in Section \ref{swallowparameter}. From here on, the physics is more transparent by forgoing this choice and leaving the location of the barrier, $\gamma$, explicit.

$\lambda_3$ is directly related to the density perturbations,
\begin{eqnarray} \label{densityswallow}
\Delta_{\mathcal{R}}^{2} &=& \frac{V(\alpha)}{12\pi^2}\lambda_3^{2}N_{e}^{4} \nonumber, \\
&=& \frac{V(\alpha)}{12\pi^2}N_e^4 \mathcal{Z}^2(\alpha-\beta)^2(\alpha-\gamma)^2\,\,.
\end{eqnarray}
%

Here, $N_e$ denotes the number of e-foldings, and $\mathcal{Z}$ is a prefactor that depends on the canonical normalization of the inflaton. Again, the density perturbations scale with the separation of the roots of $V^{\prime}(x)$.

Physically, $V(\alpha)$ corresponds to the scale of inflation and $V(\beta)$ is the cosmological constant. The energy scale of the barrier height, $V(\gamma)$ often carries other physical significance. For the example of K\"{a}hler moduli inflation discussed later, it is correlated to both the scale of SUSY breaking and decompactification of the internal space. Catastrophe Theory predicts scaling relations between them, as we now see.

For our purposes, $V(\beta)$ is vanishingly small, as is $\lambda_1$, for reasons mentioned earlier. From this, we can read off the scale of inflation
\begin{eqnarray}\label{infscale}V(\alpha) &=& V_0\frac{(-\lambda_3)}{3}(\beta-\alpha)^3,\\
&=& \frac{V_0}{3}(\beta-\alpha)^4(\gamma-\alpha).\end{eqnarray}
The scale of the barrier height is,
\begin{equation}\label{bg}V(\gamma) = V_0\frac{(-\lambda_3)}{3}\big[(\beta-\alpha)^3 - (\gamma-\alpha)^3\big].\end{equation}
In short, the energy scales of the problem scale with both $\lambda_3$ and, more importantly, the separation of the roots. A large $(\beta-\alpha)$ corresponds to a large scale of inflation. With the interpretation of $V(\gamma)$ as the scale of SUSY breaking, nearby roots imply that the scales of inflation and SUSY breaking are tied together. Thus, these scaling relations put the observations of Kallosh and Linde on precise, quantitative grounds. Additionally, we find that the scale of SUSY breaking can be separated from the inflation scale, and scales parametrically with the root separation $(\gamma-\alpha)$.

\section{Racetrack Inflation}
\subsection{Parameter Space}

In this subsection, we analyze an example of swallowtail inflation: K\"ahler (volume) modulus inflation in type IIB string theory. We deal with a single K\"ahler modulus in the KKLT paradigm, keeping multiple K\"ahler moduli for future work. First we outline the parameter space of the physical input parameters that lead to successful inflation, with zero cosmological constant. Then, we comment on the scale of inflation and its relation to supersymmetry breaking.

The main elements in the 4D $\mathcal{N}=1$ effective theory are: $(1)$ background flux induced Gukov-Vafa-Witten superpotential \cite{Gukov:1999ya}, and $(2)$ gaugino condensation on $D7$-branes \cite{Gorlich:2004qm} or Euclidean D3 instantons \cite{Witten:1996bn}, which result in non-perturbative contributions to the superpotential. These two contributions to the superpotential are sufficient to stabilize complex structure moduli and the dilaton. Moreover, the K\"ahler moduli are stabilized in an AdS vacuum. An additional contribution to the scalar potential coming from anti-D3-branes then lifts the solution to a de Sitter vacuum \cite{Kachru:2003aw}.

The tree level K\"ahler Potential is
$$K = -3\log(T + \bar{T}).$$
In \cite{Badziak:2008yg} the importance of $\alpha^{\prime}$ and $g_{s}$ corrections to $K$ for inflation was emphasized. These corrections can be parametrized schematically as,
$$K \sim -3\log\big[T + \bar{T} + \xi + \frac{a_{1}}{(T+\bar{T})^n}\big].$$
In our analysis these additional parameters $a_1$ and $\xi$ are absorbed into the Swallowtail control space. They are examples of perturbations of $V$ that do not change the domain structure of the Swallowtail Catastrophe. Thus, we no longer make reference to them explicitly. They may behave as nontrivial control parameters in multifield models.

The effective superpotential is
\begin{equation}
W = W_0 + A e^{-a T} + B e^{-b T} \,\, .
\end{equation}
The imaginary part is stabilized by the nonperturbative breaking of the axionic shift symmetry. The flux contribution to the superpotential is given by $W_0 = \int_{\rm CY_3}G_3\wedge \Omega$. Gaugino condensation of pure Super-Yang-Mills on two stacks of $D7$ branes gives the non-perturbative superpotential contributions. The factors $a$ and $b$ are of the form $2\pi/M$, where $M$ is the rank of the respective product in the gauge group. Also, $A$ and $B$ are coefficients in the non-perturbative terms whose precise dependence on moduli comes through one-loop threshold corrections to the $D7$ gauge kinetic function and higher curvature corrections on the world volume of the $D7$s. 

The $\mathcal{N}=1$ supergravity scalar potential may be computed,
\begin{equation}  \label{scalarpotential}
V = e^{K}\Big(K^{I\bar{J}}D_{I}W \overline{D_{J}W} - 3|W|^2\Big) + \frac{4C}{(T+\bar{T})^2} \,\,,
\end{equation}
where the final, ``uplifting'', term breaks supersymmetry and is treated phenomenologically, as is usual in these constructions. Such a term may come from a $\overline{D3}$ brane trapped deep at the bottom of a warped throat \cite{Kachru:2002gs,Kachru:2003aw}.

We now investigate a local range of the scalar potential which displays an inflection point and a local minimum. Clearly, the number of physical input parameters is large and treating all of them as free would entail studying a much higher order catastrophe. We therefore fix parameters that have little effect on the inflection point. Assuming that the gauge groups are fixed, we have four parameters at our disposal, $W_0, A, B, C$. While we will treat them as free, phenomenological parameters, one should not lose sight of their discrete origin in terms flux quanta. The goal, of course, is to use Catastrophe Theory to peel away model dependant details with the hope of determining what is natural about inflation in these models. 

The flux contribution $|W_0|^2$ sets the scale of the potential. In KKLT constructions, low-scale supersymmetry requires $|W_0|$ be very small \cite{Choi:2004sx}. This is usually done by assuming the magnitude of $G^{(3,0)}$ component of $G_3$ is suppressed relative to the primitive $G^{(2,1)}$ part by at least one part in $10^{10}$. We may remove $W_0$ from the set of control parameters which influence the appearance of an inflection point, keeping it as a parameter that fixes the scale of the potential. This defines a new set of control parameters

\begin{equation}
(W_0,A,B,C)\rightarrow (1,\bar{A},\bar{B},\bar{C}) = (1, \frac{A}{W_0},\frac{B}{W_0},\frac{C}{W_0^2}) \,\,.
\end{equation}
Supersymmetric AdS minima can occur, with $\bar{C}=0$, when
\begin{eqnarray}\label{klmins}
x &\sim & \frac{1}{a}\log|\bar{A}| \,\, \nonumber\; \mathrm{and} \\
x &\sim & \frac{1}{a-b}\log|\frac{\bar{A}}{\bar{B}}| \,\,.
\end{eqnarray}
In the above, $x$ is the real part of the K\"ahler modulus $T$.

Before uplifting, the racetrack scalar potential behavior can be described by the Cusp Catastrophe. Of course, supersymmetry remains unbroken in the final vacuum. While nondegenerate critical points are possible here too, such a point would have negative vacuum energy. 

The uplifting term is positive definite, and will not contribute an additional minimum. With three parameters $(\bar{A}, \bar{B}, \bar{C})$ and two minima, we are inevitably led to consider the Swallowtail Catastrophe, whose germ is $x^5$.

At this point, we can use the description of the parameter space of the Swallowtail Catastrophe, described in Section (\ref{swallowparameter}), to understand the allowed regions. Here, we have to distinguish between the physical control parameters $\bar{A}, \bar{B}, \bar{C}$ and the canonical control parameters $a,b,c$ appearing in Eq.~(\ref{st}), in terms of which the swallowtail parameter space is cleanly described. Note that mathematically, there is no preferred set of control parameters, and the physical parameters are on the same footing as the canonical ones - the potential is a real function\footnote{It was shown in \cite{Choi:2004sx} how to set all the parameters of these sorts of constructions to real numbers.}, has at most four local extrema and three parameters. It therefore \textit{must} be described by the swallowtail catastrophe near the region relevant for inflation.

\subsection{Scale of Inflation and Supersymmetry Breaking}

In this subsection, we continue our analysis of Racetrack inflation by investigating relations between the scale of inflation and the scale of supersymmetry breaking. 

The barrier height $V(\gamma)$ is roughly the scale of supersymmetry breaking \cite{Kallosh:2004yh},
\begin{equation}
V(\gamma) \, \sim \, m_{3/2} \,\,.
\end{equation}
$V(\gamma)$ vanishes if and only the uplifting coefficient $\bar{C}$ vanishes. This breaks the $A_{4}$ structure down to $A_{3}$. Eqs. (\ref{infscale}) and ~Eq.~\eqref{bg} suggest that low-scale supersymmetry entails low scale inflation.

We are investigating the behavior of the Racetrack scalar potential, $V$ in the vicinity of known extrema given in Eq.~\eqref{klmins}.

It is clear from Eq.~(\ref{infscale}) that low-scale inflation is naturally obtained when the root separation is small. The scaling relations get more complicated with increasing complexity of the Catastrophe germ, but they are predictable and they scale parametrically with the separation of the roots. However, from Eq.~(\ref{densityswallow}), it is clear that unacceptably small density perturbations are generic to this case unless $\mathcal{Z}$, which varies parametrically with the K\"ahler modulus VEV at inflation, $\alpha$ is large.
$\mathcal{Z}$ depends on the inverse K\"ahler metric; in the case of a single K\"{a}hler modulus, one obtains
\begin{equation}
\mathcal{Z} = \left(\frac{4 \alpha^2}{3} \right)^{3/2} \,\,.
\end{equation}
From Eq.~(\ref{densityswallow}), we obtain
\begin{equation}
\Delta_{\mathcal{R}}^{2} \approx \frac{16V_0 N_e^4}{243\pi^2}(\beta-\alpha)^6(\gamma-\alpha)^3\alpha^6 \,\,.
\end{equation}
Typical values are $\Delta_{\mathcal{R}}^{2} \sim 10^{-10}$ and $N_e \sim 50$. For an intermediate scale of inflation, $V_0 \sim 10^{-28}$. Then, for the correct value of density perturbations, we need
\begin{equation} \label{racetrackvalues}
\alpha \sim O(10^2-10^3) \,\,.
\end{equation}
Note that the position of the inflection point is roughly given by its value prior to uplifting
\begin{equation} \label{rootsracetrack}
\alpha \sim  \frac{1}{a}\log|\frac{A}{W_0}| \,\,.
\end{equation}
We have $V_0 \sim |W_0|^2 \, \Rightarrow \, W_0 \sim 10^{-14}$. Thus, Eq.~(\ref{rootsracetrack}) and Eq.~(\ref{racetrackvalues}) imply that for low-scale inflation, one requires $A \sim B \sim O(e^{100}-e^{1000})$. These large values naturally appear with magnetic flux threading the 4-cycles which the $D7$ branes wrap. The gauge kinetic function on the $D7$ branes is shifted by a piece $f_{\Sigma}$ that depends on the magnetic flux and the dilaton VEV \cite{Allahverdi:2009rm}.

The struggle pointed out by Kallosh and Linde \cite{Kallosh:2004yh} was that the ``uplifting'' required by large scale inflation typically destabilize the potential. For nearby roots, one can obtain high scale inflation and low scale supersymmetry, without destabilization, by supplementing the model with additional constructions \cite{He:2010uk}. Another route is to attempt the separation within the LARGE volume scenario \cite{BlancoPillado:2006he,Conlon:2007gk,Conlon:2008cj}. This work obtained a hierarchy between inflation and SUSY breaking through a hierarchy in field space between the modulus VEV where inflation occurs, and the position of the final vacuum\footnote{The basic idea is that if some amount of background sources like radiation is present, an attractor solution to the evolution equations can guide the field to the global minimum at $x=\beta$ without overshooting, even when the extrema are widely separated.}. From Eq.~\eqref{bg} it is clear that that $(\beta-\alpha)$ controls the scale of inflation and $(\gamma-\beta)$ controls the scale of SUSY breaking. Thus, any intermediate scenario is a priori possible. 

Three lessons come from the analysis of the inflection point. First, small root separation leads to similarity between the scale of inflation and SUSY breaking. This was seen in the example of volume modulus racetrack inflation. Second, if both SUSY breaking and inflation are near the intermediate scale $10^{-7}$, then the inflection point needs to be located around $\alpha \sim O(10^2-10^3)$. Third, this implies that scale of inflation can be much larger than the scale of SUSY breaking for $\alpha\ll \beta$ and $\beta\sim\gamma$.

\subsection{Racetrack Inflation: Numerical Results}

We now investigate the full scalar potential, Eq.~\eqref{scalarpotential} to demonstrate these ideas explicitly. After making our parametrization precise, we discuss the general general features and display the results in the plots below.

Low scale inflation requires large $\big<x\big>$ and small root separations $\alpha-\beta$. One way to achieve this is involves turning on magnetic flux on the 4-cycles with the $D7$ branes wrap. The flux energy exponentially expands the internal volume, for details see \cite{Allahverdi:2009rm}.

We parametrize the magnetic flux by $f_{\Sigma}$, and take
$$A,B\rightarrow e^{af_{\Sigma}}A,e^{bf_{\Sigma}}B.$$
It is easy to see that Eq.~\eqref{klmins} cause a shift $x\rightarrow x + f_{\Sigma}$. This construction allows us to absorb the entire effect of the parameter $A$ into $f_{\Sigma}$.

Note that following the discussion after Eq.~\eqref{rootsracetrack}, it should be clear that $f_{\Sigma}$ is fixed by $\alpha$ and the scale of inflation given by $W_0$; specifically, the relation is
\be \label{magflux}
f_{\Sigma} \,\, \sim \,\, \alpha \, + \, \frac{1}{a}\log W_0 \,\,.
\ee

We may think of the family of potentials with the control parameters $f,B,C$ as a bundle over the curve given by $V$. Because we are interested in the local behavior of such families of potentials, one should assign the swallowtail parameters, $a,b,c$ as coordinates describing a section near a local trivialization of the bundle around some specific point. In other words, one should be careful to avoid large excursions in the space of Swallowtail control parameters.

Concretely, we consider the fifth order Taylor polynomial as our Swallowtail function. To that end, we expand $V$ around $\big<x\big>$,
$$ V(x) = \sum_{n=0}^{5} a_n (x-\big<x\big>)^n +O\mathbf{(}(x-\big<x\big>)^6\mathbf{)}.$$
The point $\big<x\big>$ should be fairly close to the critical point. A good starting point is to construct $V$ around a known racetrack minimum, Eq.~\eqref{klmins}.

Restricting to the fifth Taylor polynomial (Jet space) is the simplest diffeomorphism consistent with the Swallowtail behavior, but has limited validity. A useful way to parametrize the faithfulness of this mapping comes from looking at $a_5$ as a function of the Racetrack control parameters. Since the germ of $V$ should be $-\frac{x^5}{5}$, we expect $a_5<0$. One must be careful then to avoid portions of parameter space where $a_5$ approaches zero\footnote{Although this can be often remedied by a simple diffeomorphism: shift $x$. A more complete treatment is discussed in Appendix \ref{appa}.}. 

A quick change of variable, 
\begin{equation}\label{tay}y = \frac{(x-\big<x\big>)}{(5a_5)^{1/5}} - \frac{a_4}{5 a_5 },\end{equation}
gives the canonical form of the swallowtail potential, parity adjusted to agree with the Racetrack,
$$V = -\frac{1}{5}y^5 - \frac{1}{3}ay^3 + \frac{1}{2}b y^2 - cy + d.$$
where
\begin{eqnarray*}
a &=& \frac{5 a_5 a_3 - 2 a_4^2}{(5 a_5)^{8/5}},\\
b &=& -\frac{25 a_2 a_5^2 -15 a_3a_4a_5 + 4a_4^3}{(5a_5)^{12/5}},\\
c &=& \frac{125 a_1a_5^3 - 50 a_2 a_4 a_5^2 + 15 a_3 a_4^2a_5 - 3a_4^4}{(5a_5)^{16/5}},\\
d &=& a_0 + \frac{a_4^5 - 25a_3 a_4^3a_5 + 125 a_2 a_4^2 a_5^2 - 625 a_1 a_4 a_5^3}{(5a_5)^{4}}.
\end{eqnarray*}
We are then free to compare $b$ and $c$ to the domain structure of the Swallowtail parameter space. From $a$,$b$ and $c$, one can use \eqref{dj} and \eqref{stcon} compute the roots $\alpha$,$\beta$ and $\gamma$.

We investigated solutions near $B=-0.0035$, $C= \frac{27}{8}10^{-3}W_{0}^{2}$ and $f_{\Sigma}=300$. To fix the overall scale of the potential, we chose $W_0=2\times 10^{-10}$.

\begin{figure}[!h]\centering
\includegraphics[width=5in]{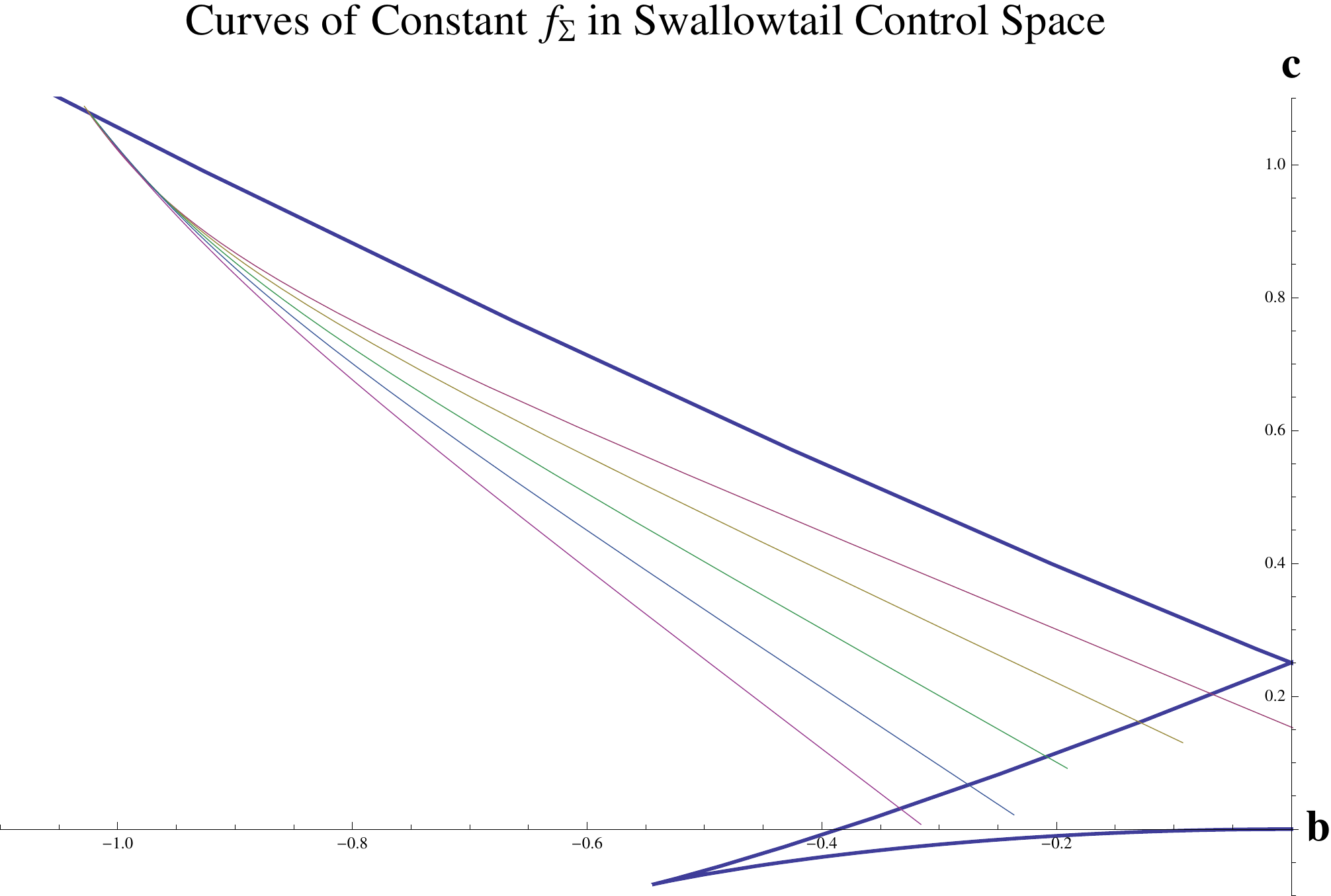}\caption{Curves of constant $B$ parameter are plotted in the Swallowtail parameter space. Here the magnetic flux parameter is fixed at 300. As $B$ is increased, the curves parametrized by $C$ move to the right. Eventually they will pass the center of the triangle, where no stable vacua exist. As $B$ decreases, eventually it will run into a cusp, requiring less and and less uplifting to reach the inflection point.}
\label{num1}
\end{figure}

Fig \ref{num1} plots curves of various $B$ parametrized by $C$ with $f_{\Sigma}=300$. Greater $B$ shifts the curves to the left. Inflation, which may occur near the triangular boundary, requires less uplifting parameter $C$. It generically corresponds to the decreasing separation of roots. Thus, $B$ is bounded from above by the left vertex of the triangle: the Cusp point. Here $\alpha=\beta$. $B$ is bounded from below by the triangle's apex, which corresponds to the loss of a stabilized minima. In terms of roots, $\beta=\gamma$. This behavior matches the full Racetrack potential.  Smaller (more negative) values of $B$ do not allow for metastable vacua. The same qualitative behavior occurs for any $f_{\Sigma}$.)

\begin{figure}[!h]\centering
\includegraphics[width=5in]{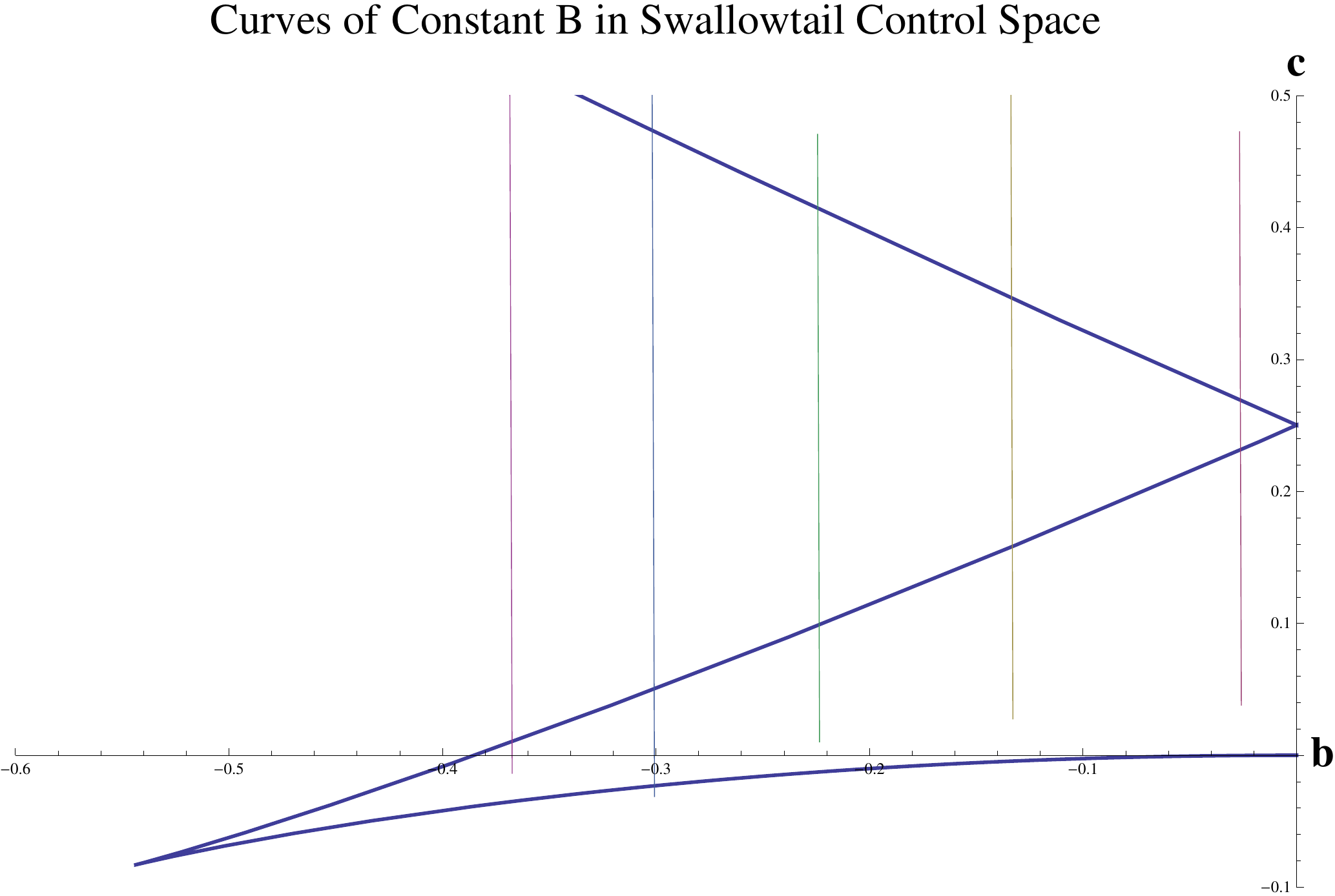}\caption{Curves of constant magnetic flux are plotted in the Swallowtail parameter space. The value of $B$ where each curve intersects the triangular region, and inflection point suitable for inflation exists on the Racetrack Potential. As magnetic flux increases, the curves come closer to the Cusp point, implying a lower ratio of inflationary and cosmological constant scales. This is all for $C= 2.7\times 10^{-22}$.}
\label{num2}
\end{figure}

Fig \ref{num2} plots curves of various $f_{\Sigma}$ parametrized by $B$ with $C= 2.7\times 10^{-22}$. Increasing the flux, $f_{\Sigma}$, pushes the curves leftward towards the Cusp point. Thus, the physical scale (greater than 30 TeV, say) of inflation bounds $f_{\Sigma}$ from above. The approximate fixed point for large $B$ is approached in the region of a single minimum. As $B$ is increased, the minima are stretched out and the barrier is flattened towards a decaying exponential behavior which is essentially independent of $f_{\Sigma}$. Of course, SUSY is broken by the nonvanishing value of $C$, so a runaway is the limiting case here. Again, the Swallowtail parametrization faithfully reconstructs the qualitative and quantitative nature of the Racetrack potential. With such analytic control of the racetrack potential, we can easily generate solutions with zero cosmological constant which was a numerical challenge in earlier works. For low scale inflation, the cosmological constant needs to be  vanishingly small so that it does not contribute to the soft supersymmetry breaking masses.

\begin{figure}[!h]\centering
\includegraphics[width=5in]{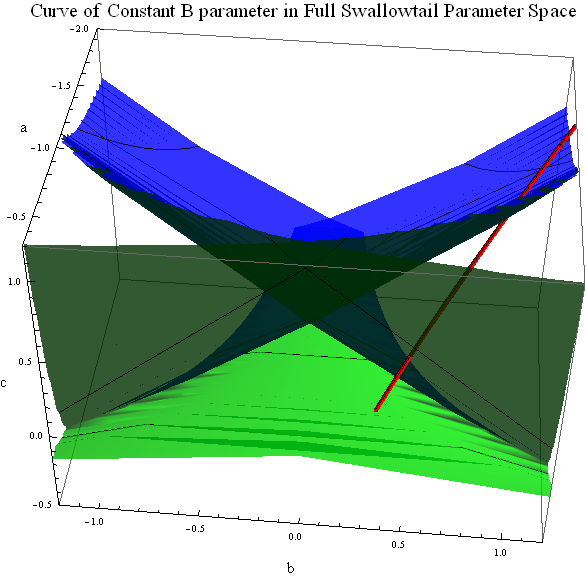}\caption{Curve of constant $C$ (thick solid line) in the three dimensional Swallowtail parameter space. As the parameter $B$ is changed, $(\beta-\alpha)$ also changes. When the curve pierces the inner surface, $B$ is suitable for inflation. As the roots approach each other the uplifting is too strong and the function is in the domain with only a single minimum. For a large change in $B$, the curve passes through the second domain wall where $V$ has runaway behavior for any $x\equiv Re[T]$.}
\label{num3}
\end{figure}

Fig \ref{num3} shows an example of a curve of constant uplift parameter $C$, in the full, three dimensional Swallowtail parameter space. In the middle domain, the function possess two minima. Changing $B$ changes the root separation. Smaller root separations mean less uplift is required to attain an inflection point. The curve pierces the domain wall at a value of $B$ which gives an inflection point suitable for inflation. Further changes brings the function into a domain of a only a single minimum. Ultimately, the curve pierces the a second domain wall. Here both the minima are lost and the potential has runaway behavior for any $x$.

\subsection{Soft Masses and Control Parameters}

When the inflaton also participates in supersymmetry breaking in the vacuum of the theory, inflationary parameters directly influence low-energy particle physics, such as the supersymmetric spectrum. In this subsection, we outline the dependence of soft masses on the position of roots in the inflaton space, as well as the root separation.

Firstly, we recap the $\mathcal{N}=1$ effective theory. The visible sector is assumed to be sequestered from the supersymmetry breaking brane, \cite{Kachru:2007xp}. The model data are,
\begin{eqnarray}
K &=& - 3 \log(T+ \bar{T})+\sum_{i}Z_i(T,\bar{T})\Phi^*_i\Phi_i, \nonumber \\
W &=& W_{0} + A e^{-af_g}+ B e^{-b f_g} +\frac{1}{6}\lambda_{ijk} \Phi_i\Phi_j\Phi_k, \nonumber \\
f_g &=&  T-f_{\Sigma} \,\, ,  \end{eqnarray}
where $\Phi_i$ are the visible sector superfields and $Z_i(T,\bar{T}) = 1/(T+\bar{T})^{n_i}$. $n_i$ is the modular weight of the superfields - either zero or one if they live on $D7$ or D3 branes, respectively. $f_{\Sigma}$ is the contribution to the gauge kinetic function coming from magnetic flux on the $D7$ branes. Recall that our effective variable is $x=\mathrm{Re}(T)$.

Since the vacuum has zero cosmological constant, we have only moduli and anomaly mediated contribution to the soft SUSY breaking masses at the GUT scale. For more details of mirage mediated SUSY breaking, see \cite{Choi:2004sx,Choi:2005uz,Endo:2005uy,Choi:2006im}. The modulus contributions to gaugino masses, trilinear couplings and sfermion masses are given by
\begin{eqnarray}
\label{tmediation} M_0 &=& F^T\partial_T\log{\rm Re}(f_g),
\nonumber \\
\tilde{m}_i^2 &=& -F^TF^{\bar{T}}\partial_T\partial_{\bar{T}}
\log(e^{-K_0/3}Z_i),
\nonumber \\
\tilde{\mathcal{A}}_{ijk} &=& F^T\partial_T\log(e^{-K_0}Z_iZ_jZ_k) \,\,.
\nonumber \\
\end{eqnarray}

Where $K_0$ is the K\"ahler potential without contributions from the superfields, $\Phi_i$. As we have discussed previously, the SUSY breaking scale is similar to the barrier height $V(\gamma)$, which in turn is similar to the inflationary scale $V(\alpha)$ for small root separation. Thus,
\begin{eqnarray}
F^2 \, &\sim & \, V(\gamma) \, \sim \, V(\alpha) \nonumber \\
&& {\rm for} \,\,\,\,\,\alpha \sim \beta \sim \gamma \,\,\,.
\end{eqnarray}
It follows from Eq.~(\ref{densityswallow}) that
\begin{equation} \label{FDelta}
F \, = \, \frac{16\pi |\Delta_{\mathcal{R}}|\alpha^3}{3N_e^2 |(\beta-\alpha)(\gamma-\alpha)|}\;.
\end{equation}

It is instructive to see what this means for the modulus contribution to the GUT scale gaugino mass, for example. Using Eq.~(\ref{FDelta}) and the expression for the gauge kinetic function, one obtains
\begin{equation}
M_0 \, = \, \frac{8\pi |\Delta_{\mathcal{R}}|\alpha^3}{3N_e^2 |(\beta-\alpha)(\gamma-\alpha)|(\beta-f_{\Sigma})}  \,\,.
\end{equation}
Note that we have used $x= \beta$ as the location where the gauge kinetic function is calculated, giving $f_g = \beta - f_{\Sigma}$. 

The anomaly contribution to the gaugino mass at the GUT scale is given by
\begin{equation}
M_{\rm anomaly} \, = \, \frac{m_{3/2}}{16\pi^2} b g^2 \,\,,
\end{equation}
where $b$ is the beta function coefficient for the relevant gaugino, and $g$ is the gauge coupling. The gravitino mass and the modulus contribution $M_0$ are related by a parameter $\alpha_{\rm mir}$ that controls the relative strengths of modulus and anomaly contributions
\begin{equation}
\alpha_{\rm mir} \, = \, \frac{m_{3/2}}{M_0 \log (M_p/m_{3/2})} \,\,.
\end{equation}
Finally, $\alpha_{\rm mir}$ can be obtained as a function of the position of the inflaton \cite{Allahverdi:2009rm}
\begin{eqnarray} \label{miragebeta}
\alpha_{\rm mir} &=& \frac{\beta}{32}\left(1-\frac{f_{\Sigma}}{\beta} \right), \; \nonumber \\
&=& \frac{\beta-f_{\Sigma}}{32}\; .
\end{eqnarray}
Putting everything together, we obtain that the total GUT scale gaugino mass is given by
\begin{equation}
M_{\rm total} \, = \, \frac{8\pi |\Delta_{\mathcal{R}}|\alpha^3}{3N_e^2 |(\beta-\alpha)(\gamma-\alpha)|(\beta-f_{\Sigma})}\left(1 + \frac{bg^2}{16\pi^2}\alpha_{\rm mir} \log (M_p/m_{3/2})\right) \,\,.
\end{equation}

First, we can fix the scale $M_0 \sim 10^{-4} - 10^{-2}$, which follows from the value of $\Delta_{\mathcal{R}} \sim 10^{-5}$ and Eq.~(\ref{racetrackvalues}). This leads to gaugino masses in the $100-1000$ GeV range after RG running. The precise value would depend on the exact separation of roots. Second, the relative importance of modulus versus anomaly mediation can be controlled by the root position $\beta$, as is clear from Eq.~\eqref{miragebeta}. Note that as mentioned before, $f_{\Sigma}$ is not an independent parameter but is given by Eq.~\eqref{magflux}.

This suggests that if we can identify $\alpha_{\rm mir}$ the ratio of the mirage to anomaly mediation contribution and $M_0$ at the LHC~\cite{kechen},  we can determine  two differences among the three roots, $\alpha$, $\beta$ and $\gamma$. Cosmology data may determine the scale of inflation, and therefore combining the data from the LHC and cosmology, we may be able to identify the nature of the potential responsible for inflation and SUSY breaking.

\section{Conclusion}

Many attempts to incorporate inflation within the framework of particle physics and string theory involve inflection points. While many models exist with markedly different motivations, we have taken the point of view that they are all related to universal archetypes given by Singularity Theory, and tried to extract physical information in a model-independent way. Our approach gives 

\textit{(i)} A neat classification of the kinds of local potentials that one can obtain for a given number of physical input parameters affecting inflation. 

\textit{(ii)} A full description of the parameter space where one can expect inflationary solutions. The space of couplings or control parameters is divided into domains with different qualitative behaviors of the scalar potential. Near a domain wall, a function possesses an inflection point suitable for inflation. Perturbations away from the degenerate critical point determine the number of e-foldings of inflation. These domain walls depend on the separation of the roots of $V^{\prime}(x)$ in a precise way. These features are stable against perturbations as long as a domain wall is not crossed. 

As mentioned in the text, the number of e-foldings $N_e$ parametrizes the unfolding of the singularities associated with degenerate critical points. To get a physically reasonable number e-foldings, the parameters of the potential must be very close to a domain wall. Universal behavior in Warped D-Brane inflation was characterized in terms of e-foldings in \cite{Agarwal:2011wm}. It is possible to generalize these results to all models using inflection points, and we have found preliminary evidence in agreement with \cite{Agarwal:2011wm}, which will be presented in a later publication.

\textit{(iii)} The fact that domain walls depend on the separation of the roots of $V^{\prime}(x)$ in a precise way turns into scaling relations for physical observables such as density perturbations and the energy scales of inflation and SUSY breaking. For low-scale inflation with metastable vacua, this analysis quantitatively explains Kallosh and Linde's bound between the inflationary and SUSY breaking scales. Low-scale SUSY breaking with high-scale inflation may be achieved with large root separation, regardless of the origin of the scalar potential. The scaling relations even appear in computations of the soft SUSY breaking masses. Successful measurement of SUSY masses at colliders will allow correlations between particle physics data and cosmology data through archetype models such as the Swallowtail. It could also rule out inflection point inflation altogether.

In short, we have taken inflection point inflation (like phase transitions) as an application of Catastrophe Theory. The application gives both qualitative and quantitative information. 

A singularity theoretic approach to inflation may also be useful for understanding the behavior of potentials as the VEV of the inflaton approaches the Planck Scale. In this limit, new, higher dimensional operators increase the dimension of the space of control parameters. Since lower order catastrophes nest inside higher order ones, inflation is still viable, but the control parameter space must be appropriately analyzed. It may be possible, then, to organize the Planck suppressed operators and make contact with particle physics. \cite{add} is a first step in this direction.

As previously mentioned, $N_e$ parametrizes the unfolding of the singularities associated with degenerate critical points. To get a physically reasonable number e-foldings, the potential's parameters must be very close to a domain wall. Universal behavior in Warped D-Brane inflation was characterized in terms of e-foldings in \cite{Agarwal:2011wm}. It is possible to generalize these results to all models using inflection points, and we have found preliminary evidence in agreement with \cite{Agarwal:2011wm}, which will be presented in a later publication.

This work has focused on $A$-type singularities. Multi-field inflation brings about the possibility of singularities of $D$- or $E$-type. Indeed, searches for multi-field models of K\"ahler Moduli inflation are numerically challenging. Yet such models seem to naturally incorporate $D$-type singularities. Similar techniques to those in this paper are readily available, and may have interesting consequences for the study of non-gaussianities in the power spectrum. We shall report on these details in future work.

\section{Acknowledgements}
This work is supported in parts by the DOE grant DE-FG02-95ER40917. We thank R. Allahverdi for useful discussions and the participants of the Aspects of Inflation workshop at the Mitchell Institute for stimulating questions.

\appendix
\section{Critical Points and Symmetry Breaking}\label{appa}
\subsection{Symmetries of Critical Points}
Catastrophes of a single variable stem from the germ, $x^{n}/n$. Under Arnold's classification \cite{arnold2}, it is an $A_{n-1}$ critical point. The fully perturbed germ is then 
\begin{equation}\label{an}f_{n}(x) = \frac{1}{n}x^{n} + \sum_{m=0}^{n-2}a_{m}x^{m}.\end{equation}
One way to understand the origin of the $A_{n-1}$ symmetry of the critical point is to interpret $f_{n}$ as the characteristic polynomial of some $n\times n$ matrix, $M$, in the adjoint representation of $A_{n-1}$ (that is, $\mathsf{SU}(n)$). The control parameters $a_{i}$ are then specific sums of products of traces of products of M, as one might be familiar with from studying the Cayley-Hamilton Theorem. For example,
$$a_{n-2} = \frac{1}{2}\big[(\mathrm{tr}M)^2 - \mathrm{tr}(M^2)\big].$$
Other important examples are $a_0 = \det M$, and $a_{n-1}=-\mathrm{tr}M=0$. The latter is a generic property of ``Gell-Mann'' matrices of $A_{n-1}$. Notice that these coefficients, by the cyclic property of trace, $a_{i}$ are manifestly invariant under a $A_{n-1}$ group action.
$$M\rightarrow g^{\dagger}Mg,$$
for $A_{n-1}$ element $g$.
Generally, $M$ is a vector in the complexified Lie Algebra of $A_{n-1}$. The result of this analysis is that the control parameters of Eq.~\eqref{an} are maps from this linear space to $\mathbb{C}$. Typically we restrict to $a_{i}\in\mathbb{R}$, but some nontrivial phases may be important in certain models of inflation, and as such we get the added bonus of being able to understand them. In either case, the complexification is still important, for if 
$$M = \sum_{j=1}^{n-1}c_{j}\lambda_{j},$$
with $\lambda_{i}$ being the Gell-Mann matrix associated with a basis element of the Cartan subalgebra of $A_{n-1}$, then 
$$a_{n-2} = -\sum_{j=1}^{n-1} c_{j}^2.$$
That $c_{j}\in\mathbb{C}$ is important to ensure the possibility that $a_{n-2}$ can be either positive or negative. Aside from filling the entire Catastrophe parameter space, $a_{n-2}=0$ is generically located on a domain wall. The upshot all this is that the domain structure of control parameter space maps nicely into the Weyl chambers of the weight space of $A_{n-1}$.

\subsection{Breaking Symmetries of Critical Points}
Breaking a Catastrophe of type $A_{n-1}$ down to type $A_{n-2}$ is a simple matter of analyzing germs,
$$\frac{x^{n}}{n}\rightarrow \frac{x^{n}}{n}+ \frac{t}{n-1}x^{n-1}.$$
Here $t$ parametrizes the breaking. Demanding that $t\neq0$ is equivalent to demanding that the matrix $M$ have a nonvanshing trace, breaking the $A_{n-1}$ symmetry. We then find ourselves with a germ of the $A_{n-2}$ catastrophe. For the polynomials we have been considering, this merely amounts to shifting $x\rightarrow x+t$. In the case of the Cusp Catastrophe, this amounts to moving along the curve depicted in Fig~\ref{cuspfig}. For a general $n$, this breaking can be extended
$$\frac{x^{n}}{n}\rightarrow \frac{x^{n}}{n}+ \frac{t}{n-1}x^{n-1} + \frac{s}{n-2}x^{n-2} + \hdots + \frac{z}{3} x^3.$$
Any further breaking will completely morsify the function to which this Catastrophe belongs. Tuning the parameters away from zero effectively ``blows up'' the various spheres in parameter space, circles in this case, associated with the mapping singularity. 

The careful reader may complain that we have simply renamed the control parameters. For the case of a polynomial functions, this is true, modulo the shift. It is in this sense that polynomials are the archetypes for more complicated functions, and our philosophy is that inflation need only be understood in terms of said archetypes. The point is that this ``Catastrophic'' behavior exists so long as the function is smooth in the relevant domain. Put another way, a simple Taylor expansion is not always the appropriate diffeomorphism to analyze this behavior. 

Concretely, let $V(x)$ be a function whose parameter space allows for a quartically degenerate critical point $-$ Swallowtail behavior. As in Eq.~\eqref{heart}, we have
\begin{equation}\label{apdst}\frac{dV}{dx} = (x-\alpha)(x-\beta)(x-\gamma)(x-\delta)f(x),\end{equation}
where $f(x)$ is nonvanshing for any $x$. If each root is different, we have a morse function with four local extrema. If $\alpha=\delta$, say, we'd have a doubly degenerate critical point. A typical flux compactification scalar potential may have $f(x) \sim e^{-x}/x^n$, which means that $V$ is certainly not a polynomial!

If the Swallowtail parameters are quite small, a fifth order Taylor polynomial will describe the functional behavior quite well. As the roots begin to separate, the Taylor series will loose fidelity. Despite this, Eq.~\ref{apdst} still guarantees the structural behavior. The domain walls in parameter space still exist, but the mapping from control parameter space to the simple Swallowtail model becomes complicated, perhaps intractably so.

Physically, however, this shouldn't concern us. What matters is simply that a doubly degenerate critical point, a local minimum and a barrier to runaway \textit{exist}. From this perspective, we can simply look for the $A_{2}$ catastrophe relevant for inflation. Let $y=x-\alpha$. The fifth order Taylor polynomial becomes
$$V(x+\alpha)\approx V_0(\frac{1}{5}y^5 + \frac{t}{4}y^4 + \frac{s}{3}y^3 + \frac{\lambda_2}{2}y^2 + \lambda_1 y) + V(\alpha).$$
The parameters $t$ and $s$ now contain the data about the barrier and metastable vacuum. If we choose $\alpha$ to be the inflection point, then,
$$V(x+\alpha) = V_0(\frac{s}{3}y^3 + \lambda_1 y) + V(\alpha) + O(y^4),$$
which is is all the data we need to investigate the number of e-foldings, as we now demonstrate. $\lambda_1$ serves as the slow roll parameter $\epsilon$. Because of the small field excursion, the number of e-foldings will be given very accurately by 
$$N_e = \int_{0}^{y_{\rm end}}\frac{dy}{s y^2 + \lambda_1}\propto 1/\lambda_1.$$
The rest of the physical properties follow from similar considerations. 

In all, a singularity theoretic approach to studying inflation is particularly well suited to embedding it within a larger physical theory. Existence of inflation depends on existence of domain walls in allows regions of parameter space. Relationships between inflation and other phenomena, like SUSY breaking are illuminated. This behavior is necessarily universal.

\vskip 0.2in

\bibliography{catbib}{}
\bibliographystyle{plain}

\end{document}